\documentclass[12pt,a4paper]{article}
\usepackage{amssymb,amsmath}
\usepackage[dvips]{lscape,graphicx}
\usepackage{color}
\usepackage{cite}
\usepackage{ulem}

\newcommand{\bc}{\begin{center}}
\newcommand{\ec}{\end{center}}
\newcommand{\bd}{\begin{displaymath}}
\newcommand{\ed}{\end{displaymath}}
\newcommand{\be}{\begin{equation}}
\newcommand{\ee}{\end{equation}}
\newcommand{\ba}{\begin{array}}
\newcommand{\ea}{\end{array}}
\newcommand{\bt}{\begin{tabular}}
\newcommand{\et}{\end{tabular}}

\newcommand{\ds}{\displaystyle}

\begin{document}

\title{
\textbf{Leptogenesis as an origin of dark matter and baryon asymmetries in
 the $E_6$ inspired SUSY models
\\[4mm]
}}

\date{}
\author{R.~Nevzorov\\[1mm]
\itshape{Alikhanov Institute for Theoretical and Experimental Physics,}\\[1mm]
\itshape{Moscow, 117218, Russia}\\[2mm]
}

\maketitle

\begin{abstract}
\noindent
We explore leptogenesis within the $E_6$ inspired $U(1)$ extension of the MSSM in which exact custodial symmetry forbids
tree-level flavour-changing transitions and the most dangerous baryon and lepton number violating operators. This
supersymmetric (SUSY) model involves extra exotic matter beyond the MSSM. In the simplest phenomenologically viable
scenarios the lightest exotic fermions are neutral and stable. These states should be substantially lighter than $1\,\mbox{eV}$
forming hot dark matter in the Universe. The low--energy effective Lagrangian of the SUSY model under consideration
possesses an approximate global $U(1)_E$ symmetry associated with the exotic states. The $U(1)_E$ symmetry is explicitly
broken because of the interactions between the right--handed neutrino superfields and exotic matter supermultiplets.
As a consequence the decays of the lightest right--handed neutrino/sneutrino give rise to both $U(1)_E$ and $U(1)_{B-L}$
asymmetries. When all right--handed neutrino/sneutrino are relatively light $\sim 10^6-10^7\,\mbox{GeV}$
the appropriate amount of the baryon asymmetry can be induced via these decays if the Yukawa couplings of the
lightest right--handed neutrino superfields to the exotic matter supermultiplets vary between $\sim 10^{-4}-10^{-3}$.
\end{abstract}
\thispagestyle{empty}
\vfill
\newpage
\setcounter{page}{1}

\section{Introduction}
The presence of baryon asymmetry and dark matter in the Universe clearly indicates the need for
new physics beyond  the Standard Model (SM). Over last forty years a number of new physics
mechanisms for baryogenesis were proposed including GUT baryogenesis \cite{gut-baryogen},
baryogenesis via leptogenesis \cite{Fukugita:1986hr}, the Affleck-Dine mechanism \cite{Affleck-Dine},
electroweak baryogenesis \cite{ew-baryogen}, etc\,.
Among these mechanisms thermal leptogenesis \cite{Fukugita:1986hr} is particularly attractive
because it can be naturally realised in the seesaw models \cite{see-saw} in which right--handed neutrinos
are superheavy shedding light on the origin of the mass hierarchy in the lepton sector. In these models
all three Sakharov conditions \cite{Sakharov:1967dj} are fulfilled since the seesaw mechanism
requires lepton number violation and complex neutrino Yukawa couplings can provide a source for
CP violation. In this case the lepton asymmetry is induced by the decays of the lightest right--handed
neutrino and gets partially converted into baryon asymmetry via sphaleron processes \cite{Kuzmin:1985mm}.
It was shown that the appropriate amount of baryon asymmetry in the SM and its minimal
supersymmetric (SUSY) extension (MSSM) can be generated only when the mass of the lightest
right--handed neutrino $M_1$ is larger than $10^9\,\mbox{GeV}$ \cite{lower-bound}.
In the supergravity (SUGRA) models this lower bound on $M_1$ results in the gravitino
problem \cite{gravitino-problem}. Indeed, after inflation the universe thermalizes with a reheat
temperature $T_R$. Thermal leptogenesis may take place if $T_R > M_1$. On the other hand
when $T_R\gtrsim 10^9\,\mbox{GeV}$ such a high reheating temperature leads to an overproduction
of gravitinos which tend to decay during or after Big Bang Nucleosynthesis (BBN) destroying the
agreement between the predicted and observed light element abundances.

In this context it is especially interesting to study thermal leptogenesis within well motivated SUSY extensions
of the SM that can originate from the heterotic superstring theory \cite{superstring} and/or Grand Unified
theories (GUTs) based on the $E_6$ gauge group or its subgroup. Near the GUT scale the gauge symmetry
in these models can be broken down to the SM gauge group together with an extra $U(1)'$ gauge symmetry
which is a linear combination of $U(1)_{\psi}$ and $U(1)_{\chi}$
\be
U(1)'=U(1)_{\chi}\cos\theta+U(1)_{\psi}\sin\theta\,.
\label{1}
\ee
In Eq.~(\ref{1}) the $U(1)_{\psi}$ and $U(1)_{\chi}$ symmetries are defined by: $E_6\to SO(10)\times U(1)_{\psi}$,
$SO(10)\to SU(5)\times U(1)_{\chi}$ (for a review see \cite{Langacker:2008yv, E6-review}). In these rank--5 model
all anomalies are canceled automatically if the low-energy spectrum involves complete $27$-plets. Thus in the $E_6$
inspired SUSY models the low energy matter content is extended to fill out three complete fundamental (27-dimensional)
representations of $E_6$. Each $27$--plet, referred to as $27_i$ with $i=1,2,3$, involves one generation of ordinary
matter, a SM singlet field $S_i$, that carries non--zero $U(1)'$ charge, up- and down-type Higgs--like doublets $H^{u}_{i}$
and $H^{d}_{i}$ and charged $\pm 1/3$ exotic quarks $D_i$ and $\bar{D}_i$.

Among the $E_6$ inspired SUSY models with additional $U(1)'$ symmetry there is only one combination of $U(1)_{\psi}$
and $U(1)_{\chi}$ for which right--handed neutrinos transform trivially, i.e. they remain sterile after the breakdown of the
$E_6$ symmetry. The corresponding $U(1)_{N}$ gauge symmetry is associated with the angle $\theta=\arctan\sqrt{15}$
in Eq.~(\ref{1}). Only in this Exceptional Supersymmetric Standard Model (E$_6$SSM) \cite{King:2005jy, King:2005my}
right-handed neutrinos can be rather heavy providing a mechanism for the generation of the baryon asymmetry
in the Universe via leptogenesis \cite{Hambye:2000bn, King:2008qb}. Moreover in this case the successful thermal
leptogenesis can be achieved without encountering a gravitino problem \cite{King:2008qb}.
Nevertheless the presence of the TeV scale exotic matter in the E$_6$SSM gives rise to the operators that results in
the non--diagonal flavour transitions and rapid proton decay.

Here we focus on the $U(1)_{N}$ extension of the MSSM in which a single discrete $\tilde{Z}^{H}_2$ symmetry
forbids such operators. In the simplest phenomenologically viable scenarios the lightest and next--to--lightest SUSY particles
(LSP and NLSP), which are predominantly the fermion components of the SM singlet superfields $S_i$, are stable and
tend to be considerably lighter than $1\,\mbox{eV}$. We argue that in this limit the Lagrangian of the SUSY model under
consideration possesses an extra global $U(1)_E$ symmetry associated with the exotic states. This symmetry is explicitly
broken due to the interactions of exotic states and right-handed neutrinos. As a result the decays of the lightest
right-handed neutrinos generate not only baryon and lepton asymmetries but also dark matter asymmetry. Our analysis
indicates that thermal leptogenesis may occur for $T_R\lesssim 10^{6-7}\,\mbox{GeV}$ in this case. For such a low reheating
temperature the gravitino density becomes sufficiently low \cite{Kohri:2005wn}. As a consequence the success of the BBN
is preserved.

The paper is organised as follows. In the next section we review the $U(1)_{N}$ extensions of the MSSM.
In section 3 we consider the generation of the baryon and dark matter asymmetries. Our results are summarized in section 4.

\section{The $U(1)_N$ extensions of the MSSM}
In this section, we briefly review the $E_6$ inspired SUSY models with extra $U(1)_N$ factor.
Within last ten years, several variants of the $U(1)_{N}$ extensions
of the MSSM have been proposed \cite{King:2005jy,King:2005my, Hall:2011zq,Nevzorov:2012hs, Athron:2014pua,King:2016wep}.
Supersymmetric models with an additional $U(1)_{N}$ gauge symmetry have
been studied in \cite{Ma:1995xk} in the context of non--standard
neutrino models with extra singlets, in \cite{Suematsu:1997au} from the
point of view of $Z-Z'$ mixing, in \cite{Keith:1997zb, Suematsu:1997au, Keith:1996fv}
where the neutralino sector was explored, in \cite{Hall:2009aj, Athron:2015vxg} in the context of
dark matter, in \cite{Keith:1997zb, King:2007uj} where the
renormalisation group (RG) flow of couplings was examined and in
\cite{Suematsu:1994qm, Keith:1997zb, Daikoku:2000ep} where the electroweak symmetry breaking was
studied. Recently, the RG flow of the Yukawa couplings and the theoretical upper bound
on the lightest Higgs boson mass were explored in the vicinity of the quasi--fixed point
\cite{Nevzorov:2013ixa} that appears as a result of the intersection of the
invariant and quasi--fixed lines \cite{Nevzorov:2001vj}.
The presence of a $Z'$ boson and additional exotic matter predicted by
these models provides distinctive LHC signatures which were analysed in \cite{King:2005jy, King:2005my, Accomando:2006ga, Athron:2011wu},
as well as results in non-standard Higgs decays \cite{Nevzorov:2013tta,Hall:2010ix, Athron:2014pua}.
Within the constrained version of the E$_6$SSM the particle spectrum has been examined
in \cite{Athron:2015vxg, Athron:2011wu, Athron:2012sq}, including the effects of threshold corrections from heavy states \cite{Athron:2012pw}.
The renormalisation of the vacuum expectation values (VEVs) and the fine tuning in the E$_6$SSM were considered
in \cite{Sperling:2013eva} and \cite{Athron:2013ipa} respectively.

In order to suppress non--diagonal flavour transitions in the E$_6$SSM an approximate $Z^{H}_2$ symmetry, under which
all superfields except one pair of $H^{d}_{i}$ and $H^{u}_{i}$ (say $H_d\equiv H^{d}_{3}$ and $H_u\equiv H^{u}_{3}$)
and one SM-type singlet field ($S\equiv S_3$) are odd \cite{King:2005jy, King:2005my}, can be imposed.
The most dangerous baryon and lepton number violating operators, that give rise to rapid proton decay, can be forbidden by
either a $Z_2^L$ symmetry, under which all superfields except leptons are even, or a $Z_2^B$ discrete symmetry which implies
that the exotic quark and lepton superfields are odd whereas the others remain even.
The discrete symmetries $Z^{H}_2$, $Z_2^L$ and  $Z_2^B$ do not commute with $E_6$ since different components
of $27$--plets transform differently under these symmetries.  The imposition of such symmetries to ameliorate
phenomenological problems is an undesirable feature of the models under consideration.

In this article we consider the $U(1)_N$ extension
of the MSSM in which a single discrete $\tilde{Z}^{H}_2$ symmetry simultaneously forbids the tree--level
flavour--changing transitions and the most dangerous baryon and lepton number violating operators. This SUSY model (SE$_6$SSM)
imply that near the GUT scale $E_6$ or its subgroup is broken down to $SU(3)_C\times SU(2)_W\times U(1)_Y\times U(1)_{N}\times Z_{2}^{M}$,
where $Z_{2}^{M}=(-1)^{3(B-L)}$ is a matter parity \cite{Nevzorov:2012hs}. Below the GUT scale the particle content
of the SE$_6$SSM includes three $27_i$--plets and a set of $M_{l}$ and $\overline{M}_l$ supermultiplets from $27'_l$ and $\overline{27'}_l$.
All superfields, that stem from complete $27_i$--plets, are odd while all supermultiplets $M_{l}$ are even under the $\tilde{Z}^{H}_2$
symmetry. The set of $M_{l}$ can be used for the breakdown of gauge symmetry and therefore should involve $H_u$, $H_d$ and $S$.
Also in the simplest case the set of supermultiplets $M_{l}$ has to include a lepton $SU(2)_W$ doublet $L_4$ to allow the lightest exotic quarks
to decay \cite{Nevzorov:2012hs}. In principle, the supermultiplets $\overline{M}_l$ can be either odd or even under the
$\tilde{Z}^{H}_2$ symmetry. In the SE$_6$SSM $\overline{S}$ and $\overline{L}_4$ are even whereas $\overline{H}_u$ and
$\overline{H}_d$ are odd under $\tilde{Z}^{H}_2$.

The $\tilde{Z}^{H}_2$ even supermultiplets $H_u$, $H_d$, $S$ and $\overline{S}$
survive to low energies and can acquire VEVs at the TeV scale breaking $SU(2)_W\times U(1)_Y\times U(1)_{N}$ gauge symmetry.
Since $S$ and $\overline{S}$ have opposite $U(1)_N$ charges the $D$-term contribution to the scalar potential may force the minimum
of this potential to be along the $D$-flat direction \cite{Kolda:1995iw}. However in such scalar potential there is a run--away direction
$\langle S \rangle = \langle \overline{S} \rangle \to \infty$. To stabilize the run-away direction we assume that in addition to $H_u$, $H_d$, $S$, $L_4$,
$\overline{S}$ and $\overline{L}_4$ the set of the $\tilde{Z}^{H}_2$-even supermultiplets involves the SM--singlet superfield $\phi$
that does not participate in the gauge interactions. It is expected that the $\tilde{Z}^{H}_2$-odd supermultiplets $\overline{H}_u$ and $\overline{H}_d$
get combined with the superposition of the corresponding components from $27_i$ forming vectorlike states with masses of order of $M_X$.
On the other hand the supermultiplets $L_4$ and $\overline{L}_4$ form TeV scale vectorlike states to render the lightest exotic quarks unstable.
The exotic quarks are leptoquarks in this case\cite{Nevzorov:2012hs}. Thus below the GUT scale the low-energy matter content in the
SE$_6$SSM involves
\be
\ba{c}
(Q_i,\,u^c_i,\,d^c_i,\,L_i,\,e^c_i,\,N^c_i)
+(D_i,\,\bar{D}_i)+(S_{i})+(H^u_{\alpha})+(H^d_{\alpha})\\[2mm]
+L_4+\overline{L}_4+S+\overline{S}+H_u+H_d+\phi\,,
\ea
\label{2}
\ee
where $\alpha=1,2$ and $i=1,2,3$. In Eq.~(\ref{2}) the left-handed quark and lepton doublets, the right-handed up- and down-type quarks,
the right-handed charged leptons and neutrinos are denoted by $Q_i$ and $L_i$,  $u_i^c$ and $d_i^c$, $e_i^c$ and $N^c_i$ respectively.
The gauge group and matter content of the SE$_6$SSM can originate from the 5D and 6D orbifold GUT models in which the splitting of
GUT multiplets can be naturally achieved \cite{Nevzorov:2012hs}. Because in the SE$_6$SSM extra matter beyond the MSSM fill in complete
$SU(5)$ representations the gauge coupling unification in this SUSY model can be achieved for any phenomenologically acceptable value of
$\alpha_3(M_Z)$, consistent with its central measured low energy value \cite{King:2007uj, Nevzorov:2012hs}.

The most general renormalisable superpotential which is allowed by the $\tilde{Z}^{H}_2$, $Z_{2}^{M}$ and
$SU(3)\times SU(2)_W\times U(1)_{Y}\times U(1)_{N}$ symmetries can be written in the following form \cite{Athron:2014pua}:
\be
\ba{rcl}
W &=& \lambda S (H_u H_d) - \sigma \phi S \overline{S} + \dfrac{\kappa}{3}\phi^3+\dfrac{\mu}{2}\phi^2+\Lambda\phi
+ \mu_L L_4\overline{L}_4+ \tilde{\sigma} \phi L_4\overline{L}_4\\[2mm]
&+& \lambda_{\alpha\beta} S (H^d_{\alpha} H^u_{\beta})+ \kappa_{ij} S (D_{i} \overline{D}_{j}) + \tilde{f}_{i\alpha} S_{i} (H^d_{\alpha}
H_u) + f_{i\alpha} S_{i} (H_d H^u_{\alpha}) \\[2mm]
&+& g^D_{ij} (Q_i L_4) \overline{D}_j+ h^E_{i\alpha} e^c_{i} (H^d_{\alpha} L_4)+ W_N + W_{\rm MSSM}(\mu=0)\,, \\[2mm]
W_N & = & \frac{1}{2} M_{ij} N_i^cN_j^c + \tilde{h}_{ik} N_i^c (H_u L_k) + h_{i\alpha}  N_i^c (H^u_{\alpha} L_4)\,.
\ea
\label{3}
\ee
The interaction $\sigma \phi S \overline{S}$
in the superpotential Eq.~(\ref{3}) stabilizes the run-away direction. When $\sigma$ is small the superfields $\phi$,
$S$ and $\overline{S}$ tend to acquire large VEVs $\sim M_S/\sigma$, where $M_S$ is a SUSY breaking scale, giving rise to an extremely
heavy $Z'$ boson  \cite{Athron:2014pua}.

In the simplest case when only $H_u$, $H_d$ and $S$ acquire non--zero VEVs ($\langle H_d \rangle = v_1/\sqrt{2}$,
$\langle H_u \rangle = v_2/\sqrt{2}$ and $\langle S \rangle = s_1/\sqrt{2}$) the Higgs sector was explored in \cite{King:2005jy}.
If CP-invariance is preserved then the Higgs spectrum involves three CP-even, one CP-odd and two charged states.
The SM singlet dominated CP-even state and the $Z'$ gauge boson have always almost the same masses.
When $\lambda < g'_1$, where $g'_1$ is the $U(1)_{N}$ gauge coupling, the SM singlet dominated Higgs boson
is the heaviest CP-even state whereas the rest of the Higgs spectrum is basically indistinguishable from the one in the MSSM.
If $\lambda\gtrsim g'_1$ the Higgs spectrum has a rather hierarchical structure, which is somewhat similar to the one in the
NMSSM with the approximate PQ symmetry \cite{Miller:2005qua,Nevzorov:2004ge}. As a result the mass matrix of the
CP--even Higgs sector can be diagonalised using the perturbation theory \cite{Nevzorov:2004ge,Nevzorov:2001um}.
For $\lambda\gtrsim g'_1$ the MSSM--like CP-even, CP-odd and charged states are almost degenerate and lie beyond
the TeV range.

When the sector responsible for the breakdown of the $SU(2)_W\times U(1)_Y\times U(1)_{N}$ symmetry involves
five supermultiplets $H_u$, $H_d$, $S$, $\overline{S}$ and $\phi$ the Higgs sector contains five CP--even, three CP--odd
and two charged states. Again one of the SM singlet dominated CP-even states is always almost degenerate with the
$Z'$ gauge boson. If $\kappa$,  $\mu$, $\Lambda$ and $\tilde{\sigma}$ are small the SE$_6$SSM possesses an
approximate global $U(1)$ symmetry that gets broken by the VEVs of $S$, $\overline{S}$
and $\phi$ resulting in a pseudo--Nambu--Goldstone boson (pNGB) $A_1$. The presence of such pNGB state may
give rise to the non-standard decay mode of the SM--like Higgs boson $h\to A_1 A_1$ if $A_1$ is lighter than
$60\,\mbox{GeV}$ \cite{Athron:2014pua}.

\section{Generation of dark matter and baryon asymmetries}
As was mentioned in the previous section, the Lagrangian of the SE$_6$SSM is invariant under both
$Z_{2}^{M}$ and $\tilde{Z}^{H}_2$ symmetries. Because $\tilde{Z}^{H}_2 = Z_{2}^{M}\times Z_{2}^{E}$
the $Z_{2}^{E}$ symmetry is also conserved. The transformation properties of
matter multiplets under the $\tilde{Z}^{H}_2$, $Z_{2}^{M}$ and $Z_{2}^{E}$ symmetries are summarized in Table~\ref{tab1}.
The conservation of the $Z_{2}^{E}$ symmetry implies that the lightest exotic state, which is predominantly a superposition
of the fermion components of $S_i$, is stable. The fermion components of $S_i$ are the lightest SUSY particles. Indeed, using
the method proposed in \cite{Hesselbach:2007te} it was shown that these states should be lighter than $60-65$ GeV \cite{Hall:2010ix}.
Although the couplings of these states to the SM gauge bosons and fermions are rather small the lightest exotic state could account
for all or some of the observed cold dark matter density if it had a mass close to half the $Z$ mass. Nevertheless in this part of
the parameter space the SM--like Higgs boson would decay almost 100\% of the time into the fermion components of $S_i$
while all other branching ratios would be extremely suppressed. Such scenario has been already ruled out by the LHC experiments.
On the other hand if the fermion components of $S_i$ are substantially lighter than $M_Z$  the annihilation cross section for
$\mbox{LSP}+\mbox{LSP} \to \mbox{SM particles}$ becomes too small resulting in a relic density that is much larger than
its measured value.

\begin{table}[ht]
\centering
\begin{tabular}{|c|c|c|c|c|c|c|c|c|}
\hline
                   &  $27_i$                  &   $27_i$                         & $27'_{H_u}$  &$27'_{S}$& $\overline{27'}_{H_u}$  &$\overline{27'}_{S}$&$27'_{L}$                &$1$\\
                   &                               &                                      &$(27'_{H_d})$&                & $(\overline{27'}_{H_d})$&                                   &$(\overline{27'}_L)$&\\
\hline
                   &$Q_i,u^c_i,d^c_i,$ &$\overline{D}_i,D_i,$    & $H_u$             & $S$        & $\overline{H}_u$             &$\overline{S}$            &$L_4$                      &$\phi$\\
                   &$L_i,e^c_i,N^c_i$  &$H^d_{i},H^u_{i},S_i$& $(H_d)$          &               & $(\overline{H}_d)$           &                                   &$(\overline{L}_4)$  &\\
\hline
$\tilde{Z}^{H}_2$& $-$            & $-$                                &  $+$               & $+$        & $-$                                    & $+$                           &$+$                        &$+$\\
\hline
$Z_{2}^{M}$       & $-$             & $+$                              & $+$                & $+$        & $+$                                  & $+$                            &$-$                          &$+$\\
\hline
$Z_{2}^{E}$        & $+$           & $-$                                & $+$                & $+$        & $-$                                    &$+$                            &$-$                           &$+$\\
\hline
\end{tabular}
\caption{Transformation properties of different matter multiplets under the discrete symmetries $\tilde{Z}^H_2$,
  $Z_{2}^{M}$ and $Z_{2}^{E}$.}
\label{tab1}
\end{table}

The simplest phenomenologically viable scenarios imply that the fermion components of $S_i$ are considerably lighter than
$1\,\mbox{eV}$\footnote{The presence of very light neutral fermions in the particle spectrum might have interesting implications for
neutrino physics (see, for example \cite{Frere:1996gb}).}.
In this case the lightest exotic states form hot dark matter (dark radiation) in the Universe. At the same time the invariance of the Lagrangian
of the SE$_6$SSM under the $Z_{2}^{M}$ symmetry ensures that $R$-parity is also conserved and the lightest ordinary neutralino can be stable.
When the masses of the fermion components of $S_i$ are substantially smaller than $1\,\mbox{eV}$ these states give only a very minor
contribution to the dark matter density whereas the lightest ordinary neutralino may account for all or some of the observed cold dark matter
density. So light fermion components of $S_i$ do not affect BBN if $Z'$ boson is sufficiently heavy  \cite{Hall:2011zq}.

\begin{table}[ht]
\centering
\begin{tabular}{|c|c|c|c|c|c|c|}
\hline
&  $H^{u}_{\alpha}$ & $H^{d}_{\alpha}$ & $D_i$ & $\overline{D}_i$ & $L_4$ & $\overline{L}_4$ \\
\hline
$Q^{E}_i$ & +1  & -1  & +1 & -1  & +1  & -1   \\
\hline
\end{tabular}
\caption{The $U(1)_E$ charges of exotic matter supermultiplets in the SE$_6$SSM.}
\label{tab2}
\end{table}

These scenarios are realised when $\tilde{f}_{i\alpha}\sim f_{i\alpha}\lesssim 10^{-6}$. If the Yukawa couplings of the superfields $S_i$
are negligibly small then the terms $\tilde{f}_{i\alpha} S_{i} (H^d_{\alpha} H_u)$  and $f_{i\alpha} S_{i} (H_d H^u_{\alpha})$ in the
superpotential (\ref{3}) can be ignored. In this limit the low--energy effective Lagrangian of the SE$_6$SSM possesses an approximate
global $U(1)_E$ symmetry below the scale $M_1$ where $M_1$ is the mass of the lightest right--handed neutrinos.
Only exotic matter supermultiplets $H^{u}_{\alpha}$, $H^{d}_{\alpha}$, $D_i$, $\overline{D}_i$, $L_4$ and $\overline{L}_4$
transform non-trivially under this $U(1)_E$ symmetry. The $U(1)_E$ charges of the exotic matter fields are summarised in Table~\ref{tab2}.
Similarly to the $U(1)_{B-L}$ symmetry $U(1)_E$ is anomaly--free, if the scale of the $U(1)_N$ symmetry breaking is quite high (say,
larger than $M_1$), and it is explicitly broken because of the interactions of matter supermultiplets with $N_i^c$
in $W_N$. Thus, one can expect that both $U(1)_E$ and $U(1)_{B-L}$ asymmetries induced by the decays of the lightest right--handed neutrino/sneutrino
would not be washed out by any perturbative or non-perturbative effects in the limit $\tilde{f}_{i\alpha}, f_{i\alpha} \to 0$.
Moreover the sufficiently small values of the $U(1)_E$ violating Yukawa couplings, i.e $\tilde{f}_{i\alpha}, f_{i\alpha} \lesssim 10^{-7}$, should not
erase the generated $U(1)_E$ asymmetry as well \cite{Campbell:1990fa}. The non-zero values of $\tilde{f}_{i\alpha}$ and $f_{i\alpha}$
explicitly break the $U(1)_E$ symmetry to $Z_{2}^{E}$ and the lightest exotic particle that carries the $U(1)_E$ charge becomes unstable.
If this state is mostly a linear superposition of the fermion components of $H^{u}_{\alpha}$ and
$H^{d}_{\alpha}$ then it decays into the fermion components of $S_i$ and either $Z$ or $W$.
As a consequence the induced $U(1)_E$ asymmetry gets converted into the hot dark matter density.

It is worth noting that the non--zero values of $\tilde{f}_{i\alpha}$ and $f_{i\alpha}$ do not always violate
the $U(1)_E$ symmetry. For instance, the structure of the Yukawa interactions of $S_i$ can be such that two superfields $S_i$ carry
opposite $U(1)_E$ charges while the third one does not transform under $U(1)_E$. This happens, for example, when
$\tilde{f}_{1\alpha}\simeq  f_{2\alpha}\simeq \tilde{f}_{3\alpha}\simeq f_{3\alpha}\to 0$. In this limit, again, the $U(1)_E$ symmetry
remains anomaly--free below $M_1$ if the scale, where the breakdown of the $U(1)_N$ gauge symmetry takes place,
is higher than $M_1$.

Although the VEVs $\langle S \rangle \simeq \langle \overline{S}\rangle$ are allowed to be of the order of $M_1$ or even higher
in our analysis we assume that the SUSY breaking scale and the masses of all exotic states are much lower than $M_1$.
To avoid the gravitino problem we set $M_1\simeq 10^6\,\mbox{GeV}$. Because the decays of the lightest right--handed
neutrino/sneutrino into the final states with lepton number $L=\pm 1$ are kinematically allowed these processes create lepton
asymmetry in the early Universe which is controlled by the flavour dependent CP (decay) asymmetries. Due to $(B+L)$--violating
sphaleron interactions the induced lepton asymmetry gets converted into the baryon asymmetry. Assuming the type I seesaw mechanism
of neutrino mass generation one can define three decay asymmetries associated with the three lepton flavours $e,\,\mu$ and $\tau$
in the SM which are given by
\begin{equation}
\varepsilon_{1,\,\ell_k}=\dfrac{\Gamma_{N_1 \ell_{k}}-\Gamma_{N_1 \bar{\ell}_{k}}}
{\sum_{m} \left(\Gamma_{N_1 \ell_{m}}+\Gamma_{N_1 \bar{\ell}_{m}}\right)}\,.
\label{5}
\end{equation}
In Eq.~(\ref{5}) $\Gamma_{N_1 \ell_{k}}$ and $\Gamma_{N_1 \bar{\ell}_{k}}$ are partial decay widths of $N_1\to L_k+H_u$
and $N_1\to \overline{L}_k+H^{*}_u$ with $k,m=1,2,3$. At the tree level $\Gamma_{N_1 \ell_{k}}=\Gamma_{N_1 \bar{\ell}_{k}}$
and $\varepsilon_{1,\,\ell_k}=0$. The non--zero values of the CP asymmetries arise because of the interference between the tree--level
amplitudes of the $N_1$ decays and one--loop corrections to them.

In SUSY extensions of the SM the right--handed neutrinos can also decay into sleptons $\widetilde{L}_k$ and Higgsino
$\widetilde{H}_u$ giving rise to the CP asymmetries
\begin{equation}
\varepsilon_{1,\,\widetilde{\ell}_k}=\dfrac{\Gamma_{N_1 \widetilde{\ell}_{k}}-\Gamma_{N_1 \widetilde{\ell}^{*}_{k}}}
{\sum_{m} \left(\Gamma_{N_1 \widetilde{\ell}_{m}}+\Gamma_{N_1 \widetilde{\ell}^{*}_{m}}\right)}\,.
\label{6}
\end{equation}
The decays of the right--handed sneutrinos contribute to the generation of the total lepton asymmetry in SUSY models as well.
The corresponding CP asymmetries can be defined similarly to the neutrino ones
\begin{equation}
\varepsilon_{\widetilde{1},\,\ell_k}=\dfrac{\Gamma_{\widetilde{N}_1^{*} \ell_{k}}-\Gamma_{\widetilde{N}_1 \bar{\ell}_{k}}}
{\sum_{m} \left(\Gamma_{\widetilde{N}_1^{*} \ell_{m}}+\Gamma_{\widetilde{N}_1 \bar{\ell}_{m}}\right)}\,,\qquad
\varepsilon_{\widetilde{1},\,\widetilde{\ell}_k}=\dfrac{\Gamma_{\widetilde{N}_1 \widetilde{\ell}_{k}}-\Gamma_{\widetilde{N}_1^{*}
\widetilde{\ell}^{*}_{k}}}
{\sum_{m} \left(\Gamma_{\widetilde{N}_1 \widetilde{\ell}_{m}}+\Gamma_{\widetilde{N}_1^{*} \widetilde{\ell}^{*}_{m}}\right)}\,.
\label{7}
\end{equation}
When SUSY breaking scale is negligibly small as compared with $M_1$
\begin{equation}
\varepsilon_{1,\,\ell_k}=\varepsilon_{1,\,\widetilde{\ell}_k}=\varepsilon_{\widetilde{1},\,\ell_k}=
\varepsilon_{\widetilde{1},\,\widetilde{\ell}_k}\,.
\label{8}
\end{equation}
In the type I seesaw models the decay asymmetries mentioned above were calculated initially within the SM \cite{CPasym-SM} and
MSSM \cite{CPasym-SUSY}. In the early studies flavour effects were ignored (see for example \cite{Buchmuller:2004nz}).
The importance of these effects was emphasised in \cite{CPleptogen-flavour}.

In the non-minimal SUSY models in which the right--handed neutrinos can decay into a few lepton (slepton) multiplets and a few $SU(2)_W$ doublets,
that have quantum numbers of Higgs (Higgsino) fields, i.e. $H^u_k$ and $L_x$ ($\widetilde{H}^u_k$ and $\widetilde{L}_x$), the definitions of the
CP asymmetries (\ref{5}) and (\ref{6}) can be generalised in the following way  \cite{King:2008qb}
\begin{equation}
\begin{array}{rcl}
\varepsilon^{k}_{1,\,f}&=&\dfrac{\Gamma^{k}_{N_1 f}-\Gamma^{k}_{N_1 \bar{f}}}
{\sum_{m,\,f'} \left(\Gamma^{m}_{N_1 f'}+\Gamma^{m}_{N_1 \bar{f}'}\right)}\,,
\end{array}
\label{9}
\end{equation}
where $f$ and $f'$ can be either $\ell_x$ or $\widetilde{\ell}_x$ while $\bar{f}$ and $\bar{f}'$ may be either $\bar{\ell}_x$ or $\widetilde{\ell}_x^{*}$.
The superscripts $k$ and $m$ represent the components of the supermultiplets $H^{u}_{k}$ and $H^{u}_{m}$ in the final state.
In these SUSY models the definitions of the CP asymmetries associated with the decays of the lightest right--handed sneutrino
can be modified similarly to the neutrino ones. In order to obtain the appropriate expressions for $\varepsilon^{k}_{\widetilde{1},\,f}$
the right--handed neutrino field in Eqs.~(\ref{9}) should to be replaced by either $\widetilde{N}_1$ or $\widetilde{N}_1^{*}$.
In this case the relation between different types of CP asymmetries (\ref{8}) remains intact, i.e. $\varepsilon^{k}_{1,\,f}=\varepsilon^{k}_{\widetilde{1},\,f}$.

For $M_1\simeq 10^6\,\mbox{GeV}$ the Yukawa couplings of the superfield $N_1$ to the supermultiplets $H_u$ and $L_i$ in the SE$_6$SSM
should be quite small in order to reproduce the left--handed neutrino mass scale $m_{\nu}\lesssim 0.1\,\mbox{eV}$, i.e. $|\tilde{h}_{1k}|^2\ll 10^{-8}$.
If two other right-handed neutrino states are also rather light their Yukawa couplings to $H_u$ and $L_i$  tend to be very small as well.
So small Yukawa couplings and CP asymmetries associated with them can be ignored in the leading approximation.
Nevertheless the presence of exotic matter supermultiplets $H^{u}_{\alpha}$ and $L_4$ gives rise to the new channels of the decays of
the lightest right--handed neutrino and its superpartner, i.e.
\begin{equation}
N_1\to L_4 + H^u_{\alpha},\qquad N_1\to \widetilde{L}_4+\widetilde{H}^u_{\alpha},\qquad
\widetilde{N}^{*}_1\to L_4 + \widetilde{H}^{u}_{\alpha},\qquad \widetilde{N}_1\to \widetilde{L}_4+ H^u_{\alpha},
\label{10}
\end{equation}
At the tree level SUSY implies that the partial decay widths associated with the new channels (\ref{10}) are given by
\begin{equation}
\Gamma^{\alpha}_{N_1 \ell_4}+\Gamma^{\alpha}_{N_1 \bar{\ell}_4}=\Gamma^{\alpha}_{N_1 \widetilde{\ell}_4}+\Gamma^{\alpha}_{N_1 \widetilde{\ell}^{*}_4}=
\Gamma^{\alpha}_{\widetilde{N}_1^{*}\ell_4}=\Gamma^{\alpha}_{\widetilde{N}_1 \bar{\ell}_4}=\Gamma^{\alpha}_{\widetilde{N}_1 \widetilde{\ell}_4}=
\Gamma^{\alpha}_{\widetilde{N}_1^{*} \widetilde{\ell}_4^{*}}=\dfrac{|h_{1\alpha}|^2}{8\pi}\,M_1\,,
\label{11}
\end{equation}
Since the interactions between the superfields $N^c_i$ and supermultiplets $H^{u}_{\alpha}$ and $L_4$ violate not only
$U(1)_L$, which ensures the conservation of lepton number, but also $U(1)_E$ global symmetry the decay channels of the lightest right--handed neutrino/sneutrino (\ref{10})
induce both $U(1)_L$ and $U(1)_E$ asymmetries. They are determined by the same set of the CP asymmetries $\varepsilon^{\alpha}_{1,\,\ell_4}$,
$\varepsilon^{\alpha}_{1,\,\widetilde{\ell}_4}$, $\varepsilon^{\alpha}_{\widetilde{1},\,\ell_4}$ and $\varepsilon^{\alpha}_{\widetilde{1},\,\widetilde{\ell}_4}$.
Neglecting the Yukawa couplings $\tilde{h}_{ik}$ we get
\begin{equation}
\begin{array}{c}
\varepsilon^{\alpha}_{1,\,\ell_4}=\varepsilon^{\alpha}_{1,\,\widetilde{\ell}_4}=\varepsilon^{\alpha}_{\widetilde{1},\,\ell_4}=
\varepsilon^{\alpha}_{\widetilde{1},\,\widetilde{\ell}_4}=\ds\frac{1}{8\pi}\frac{\sum_{j=2,3}
\mbox{Im}\biggl[h^{*}_{1\alpha} B_{j} h_{j\alpha} \biggr]}{\sum_{\beta} |h_{1\beta}|^2}\,,\\[4mm]
B_{j}=\sum_{\beta}\left\{h^{*}_{1\beta} h_{j\beta}\, f\left(\ds\frac{M^2_j}{M_1^2}\right)+
\ds\frac{M_1}{M_j} h_{1\beta} h^{*}_{j\beta}f^S\left(\ds\frac{M^2_j}{M_1^2}\right)\right \}\,,
\end{array}
\label{12}
\end{equation}
$$
f(z)=f^{V}(z)+f^{S}(z)\,,\qquad f^S(z)=\dfrac{2\sqrt{z}}{1-z}\,,\qquad f^V(z)=-\sqrt{z}\,\ln\left(\dfrac{1+z}{z}\right)\,.
$$

From the superpotential (\ref{3}) it follows that the supermultiplets $H^{u}_{\alpha}$ can be always redefined in such a way that only one
doublet $H^{u}_{1}$ interacts with $L_4$ and $N^c_1$. Thus without loss of generality $\tilde{h}_{12}$ in $W_N$ may be set to zero in
the first approximation. If $h_{j1}=|h_{j1}| e^{i\varphi_{j1}}$ and $M_j$ are real the subset of the CP asymmetries
that determines the $U(1)_L$ and $U(1)_E$ asymmetries in this limit can be written as
\begin{equation}
\varepsilon^{1}_{1,\,\ell_4}=\varepsilon^{1}_{1,\,\widetilde{\ell}_4}=
\varepsilon^{1}_{\widetilde{1},\,\ell_4}=\varepsilon^{1}_{\widetilde{1},\,\widetilde{\ell}_4}=
\frac{1}{8\pi}\Biggl[\sum_{j=2,3} |h_{j1}|^2 f\left(\frac{M^2_j}{M_1^2}\right) \sin 2\Delta\varphi_{j1}\Biggr]\,,
\label{13}
\end{equation}
where $\Delta\varphi_{j1}=\varphi_{j1}-\varphi_{11}$\,.

The evolution of the dark matter and lepton number densities are described by the system of Boltzmann equations.
In the limit under consideration the results obtained in the SM and MSSM for the lepton and baryon
asymmetries can be easily generalised. In particular, the induced baryon asymmetry can be estimated as follows
(see~\cite{Davidson:2008bu}):
\begin{equation}
Y_{\Delta B}\sim 10^{-3} \varepsilon^{1}_{1,\,\ell_4} \eta \,,
\label{14}
\end{equation}
where $Y_{\Delta B}$ is the baryon asymmetry relative to the entropy density, i.e.
$$
Y_{\Delta B}=\dfrac{n_B-n_{\bar{B}}}{s}\biggl|_0=(8.75\pm 0.23)\times 10^{-11}\,.
$$
In Eq.~(\ref{14}) $\eta$ is an efficiency factor that varies from 0 to 1.
In the strong washout scenario the efficiency factor is given by
\begin{equation}
\begin{array}{c}
\eta \simeq H(T=M_1)/\Gamma_{1}\,,\\[3mm]
\Gamma_1 = \Gamma^{1}_{N_1 \ell_4}+\Gamma^{1}_{N_1 \bar{\ell}_4}=\dfrac{|h_{11}|^2}{8\pi}\,M_1\,,
\qquad\qquad
H=1.66 g_{*}^{1/2}\dfrac{T^2}{M_{Pl}}\,,
\end{array}
\label{15}
\end{equation}
where $H$ is the Hubble expansion rate and $g_{*}=n_b+\dfrac{7}{8}\,n_f$ is the number of relativistic
degrees of freedom in the thermal bath. Within the SM $g_{*}=106.75$  while in the SE$_6$SSM $g_{*}=360$.

\begin{figure}
\includegraphics[width=75mm,height=55mm]{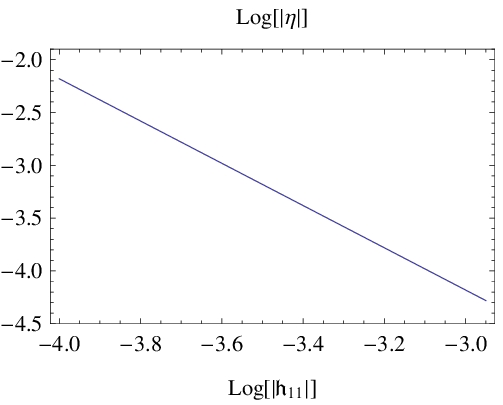}\qquad
\includegraphics[width=75mm,height=55mm]{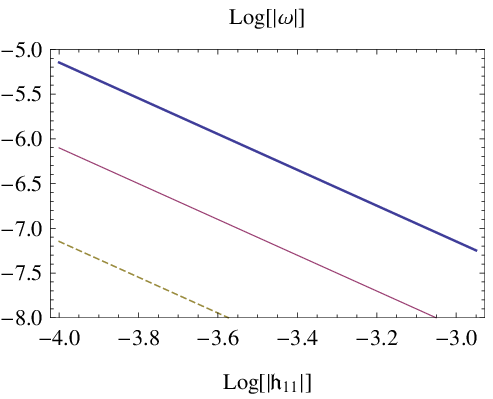}\\[2mm]
\hspace*{3.5cm}{\bf (a)}\hspace*{8cm}{\bf (b) }\\
\caption{Logarithm (base 10) of the absolute values of the efficiency factor and $\omega=\varepsilon^{1}_{1,\,\ell_4} \eta$
for $h_{i2}=h_{31}=0$, $\Delta\varphi_{21}=\pi/4$ and $M_2=10\cdot M_1$. In {\it (a)} the absolute value of
the efficiency factor is given as a function of logarithm (base 10) of $|h_{11}|$. In {\it (b)} the absolute value of
$\omega=\varepsilon^{1}_{1,\,\ell_4} \eta$ is presented as a function of logarithm (base 10) of $|h_{11}|$ for
$|h_{21}|=0.3$ (thick line), $|h_{21}|=0.1$ (solid line) and $|h_{21}|=0.03$ (dashed line).
}
\label{fig1}
\end{figure}

From Eq.~(\ref{13}) it follows that the values of the CP asymmetries are determined by the combinations of
the CP--violating phases $\Delta\varphi_{j1}$ and the absolute values of the Yukawa couplings $|h_{21}|$ and $|h_{31}|$
but do not depend on $|h_{11}|$. To simplify our numerical analysis we assume here that $|h_{31}|$ is negligibly small,
i.e. $|h_{31}|\ll |h_{21}|$, and can be ignored. We also fix $(M_2/M_1)=10$. On the other hand the efficiency factor $\eta$
is set by the lightest right--handed neutrino mass $M_1$ and $|h_{11}|$. We restrict our consideration here by the values
of $|h_{11}|$ which are considerably larger than $|\tilde{h}_{ik}|$, i.e. $|h_{11}|^2 \gtrsim 10^{-8}$, so that all Yukawa
couplings $\tilde{h}_{ik}$ can be neglected. Fig.~1a illustrates that in the strong washout scenario the efficiency factor
$\eta$ varies from $10^{-2}$ to $10^{-4}$ when $|h_{11}|$ increases from $10^{-4}$ to $10^{-3}$. The dependence
of the absolute value of $\omega=\varepsilon^{1}_{1,\,\ell_4} \eta$, that determines the induced baryon asymmetry (\ref{14}),
on $|h_{21}|$ and $|h_{11}|$ is examined in Fig.~1b. This figure demonstrates that for $\Delta\varphi_{21}=\pi/4$
and $|h_{21}|\sim 0.1$ the phenomenologically acceptable baryon density, corresponding to $\omega \sim 10^{-7}-10^{-6}$,
can be obtained if $|h_{11}|$ changes between $10^{-4}$ and $10^{-3}$. When all $U(1)_E$ violating Yukawa couplings
are very small ($\ll 10^{-7}$) the generated dark matter and baryon number densities should be of the same order of magnitude.

\section{Conclusions}
In this article we studied the generation of the baryon and dark matter asymmetries within the $E_6$ inspired SUSY model
with extra $U(1)_N$ factor (SE$_6$SSM) in which a single discrete $\tilde{Z}^{H}_2$ symmetry forbids the tree--level flavour--changing transitions
as well as the most dangerous baryon and lepton number violating operators. Only in this $E_6$ inspired $U(1)$ extension of the MSSM
the right--handed neutrinos $N^c_i$ do not participate in the gauge interactions. Therefore the decays of the heavy right--handed neutrino/sneutrino
should lead to the generation of the baryon asymmetry via leptogenesis.

To ensure anomaly cancellation the low energy matter content of the SE$_6$SSM includes three $27$ representations of $E_6$ that involve three
families of quarks and leptons, three families of exotic quarks $D_i$ and $\bar{D}_i$, three families of Higgs--like doublets $H^{d}_{i}$ and $H^{u}_{i}$
as well as three SM singlet superfields $S_i$  that carry $U(1)_{N}$ charges. In addition the particle spectrum of the SE$_6$SSM contains a pair
of $SU(2)_W$ doublets $L_4$ and $\overline{L}_4$, that allows the lightest exotic quarks to decay and facilitates gauge coupling unification,
and the SM singlet superfields $S$ and $\overline{S}$ that acquire VEVs breaking the $U(1)_N$ gauge symmetry. One pair of the Higgs--like doublet
supermultiplets $H_u$ and $H_d$ play a role of the MSSM Higgs fields which break electroweak symmetry whereas two other pairs
$H^{d}_{\alpha}$ and $H^{u}_{\alpha}$, as well as $L_4$, $\overline{L}_4$, $D_i$, $\bar{D}_i$ and $S_i$ form exotic sector. The Lagrangian
of the SE$_6$SSM is invariant with respect to $Z_{2}^{E}$ symmetry, under which all components of the exotic matter supermultiplets are odd
while all other fields are even. This discrete symmetry guarantees that the lightest exotic state, which tends to be the superposition of the fermion
components of $S_i$, is stable.

In the simplest phenomenologically viable scenarios the fermion components of $S_i$ should be significantly lighter than
$1\,\mbox{eV}$ forming hot dark matter in the Universe. These scenarios are realised if the Yukawa couplings of $S_i$ are rather small ($\lesssim 10^{-6}$).
In this limit the low--energy effective Lagrangian of the SE$_6$SSM below the scale $M_1$ possesses an approximate global $U(1)_E$ symmetry
associated with the exotic matter supermultiplets. When the scale of the $U(1)_N$ symmetry breaking is quite high ($\gtrsim M_1$) the $U(1)_E$
is anomaly--free. The interactions of  $N_i^c$ with $L_4$ and $H^{u}_{\alpha}$ break this symmetry. Thus in the SE$_6$SSM the decays of the lightest
right--handed neutrino/sneutrino induce both $U(1)_{B-L}$ and $U(1)_E$ asymmetries. It is expected that the generated $U(1)_E$ asymmetry
should not be erased if the $U(1)_E$ violating Yukawa couplings of $S_i$ are smaller than $10^{-7}$.

To avoid the gravitino problem in our analysis we focused on the scenarios with $M_1\simeq 10^6\,\mbox{GeV}$. In the case when
all right--handed neutrino/sneutrino have masses of the order of  $10^6-10^7\,\mbox{GeV}$ the Yukawa couplings of $N_i^c$ to the
left--handed lepton supermultiplets and $H_u$ tend to be rather small and can be neglected in the leading approximation. Then
the $U(1)_{B-L}$ and $U(1)_E$ asymmetries are induced because of the interactions between $N_i^c$ and the components of
the exotic matter supermultiplets $L_4$ and $H^{u}_{\alpha}$. We argued that the appropriate value of the baryon asymmetry can be
generated if the Yukawa couplings of $N_1^c$ to $L_4$ and $H^{u}_{\alpha}$ $\sim 10^{-4}-10^{-3}$.
In this case the hot dark matter and baryon number densities should be of the same order of magnitude.

\vspace{-5mm}
\section*{Acknowledgements}
\vspace{-3mm}
R.N. is grateful to M.~Binjonaid, D.~Gorbunov, V.~Novikov, O.~Kancheli, S.~F.~King, M.~M\"{u}hlleitner, M.~Sher, Yu.~A.~Simonov, D.~G.~Sutherland, 
S.~Troitsky, X.~Tata and M.~Vysotsky for  fruitful discussions.

\end{document}